\documentclass{jps-cp}
\usepackage{txfonts} 

\usepackage{graphicx,color}
\usepackage{amsmath,amssymb,bm,ascmac,calc}
\usepackage{wrapfig}

\title{$^{86}$Ni: Possible doubly-magic nucleus with neutron halo}

\author{H. \textsc{Nakada}$^{1,2}$ and K. \textsc{Ishida}$^{3}$}

\inst{
  $^{1}$Department of Physics, Graduate School of Science, Chiba University,
  Inage, Chiba 263-8522, Japan\\
  $^{2}$Research Center for Nuclear Physics, Osaka University,
  Ibaraki, Osaka 567-0047, Japan\\
  $^{3}$Department of Physics, Graduate School of Science and Engineering, Chiba University,
  Inage, Chiba 263-8522, Japan
}

\email{nakada@faculty.chiba-u.jp}

\recdate{September 8, 2026}

\abst{
  Self-consistent mean-field calculations with semi-realistic interactions
  have predicted that $N=58$ becomes a magic number near Ni,
  where the $n2s_{1/2}$ orbit is the highest occupied state.
  As the last neutrons occupy the shallowly bound $s$-orbit,
  $^{86}$Ni can be highly distinctive,
  being a doubly-magic nucleus and possessing a neutron halo simultaneously.
  With minimal ambiguity in the many-body wave function,
  it may provide a unique opportunity to investigate,
  \textit{e.g.}, reaction mechanisms in halo nuclei.
  As an example, we present differential cross sections
  for proton elastic scattering predicted
  via a folding potential without local approximations,
  and compare them with those obtained using an empirical potential.
}

\kword{New magic numbers, Neutron halo, Self-consistent mean-field calculations,
  Folding potential}

\begin{document}
\maketitle

\section{Introduction}

Extraordinary properties of nuclei far off the $\beta$ stability
have been disclosed
since the invention of radioactive nuclear beams.
Prominent examples are nuclear halos~\cite{Tan85}
and appearance of new magic numbers~\cite{Oza00}.
We have pointed out that $^{86}$Ni is a unique and intriguing nucleus
that may have both exotic properties simultaneously~\cite{Nak10,IN25},
based on the self-consistent mean-field (SCMF) calculations
with a semi-realistic interaction~\cite{Nak20}.

Neutron halos are one of the exotic nuclear properties
specific to the nuclei near the neutron drip-line~\cite{Rii94,Tan13}.
When the last neutrons have sizable $s$- or $p$-wave components
near the drip line,
the low centrifugal barrier allows them to distribute broadly
and form a halo.
While nuclear deformation helps the last neutron
acquire an $s$ or $p$ component~\cite{Ham17,NT18},
halos should be rather exceptional at magicity,
because the nuclei remain spherical.
It is noted that the quenching of pair correlations at magicity
assists formation of halos for even-$N$ nuclei
by suppressing the pairing anti-halo effect~\cite{BDP00}.

Once believed to be immutable,
we now know that nuclear magic numbers may vary
depending on the proton ($Z$) and neutron ($N$) numbers.
The tensor force plays a significant role
in this $Z,N$-dependence~\cite{Ots05,NS14}.
Since it provided the shell structure from scratch,
the SCMF framework is suitable for investigating
the appearance and disappearance of magic numbers.
In particular, the semi-realistic interaction M3Y-P6~\cite{Nak13},
whose tensor channel originates from the $G$-matrix
and is therefore considered realistic,
is advantageous in describing the nuclear shell structure,
predicting magic numbers compatible
with almost all available data~\cite{NS14,SNM16,MN18}.

In a study on shell structure in the Ca-Ni region
via SCMF calculations with a semi-realistic interaction,
one of the authors (H.N.) predicted $N=58$ to be magic at Ni~\cite{Nak10},
as the $n2s_{1/2}$ orbit comes down while $n0g_{7/2}$ goes up.
We reinvestigate the $N=58$ magicity around Ni using M3Y-P6.
Beyond its uniqueness,
the coexistence of magicity and a halo provides an indispensable opportunity.
Owing to the doubly magic nature,
the many-body content of the nuclear wave function has minimal ambiguity.
Thereby, $^{86}$Ni may supply a good testing ground for the applicability
of theories to halo nuclei,
\textit{e.g.}, reaction theories.
The optical potential is fundamental to many nuclear reaction theories.
Empirical forms of the nucleon-nucleus ($N$-$A$) optical potential
have been explored using a large amount of available data~\cite{CH89,KD03}.
However, they have been fitted to nuclei having normal density distributions,
and their applicability to halo nuclei is not obvious.
The folding potential~\cite{SL79} provides optical potentials
from the nucleonic interaction,
to which the Brieva-Rook local approximation (BR-LA)~\cite{BR-LA}
has customarily been applied.
The validity of the BR-LA may also be questioned.
Taking $^{86}$Ni as an example,
we have examined these points~\cite{IN25}.

\section{SCMF approach to $^{86}$Ni with semi-realistic interaction}

\begin{figure}[tbh]
\begin{center}
\includegraphics[scale=0.45]{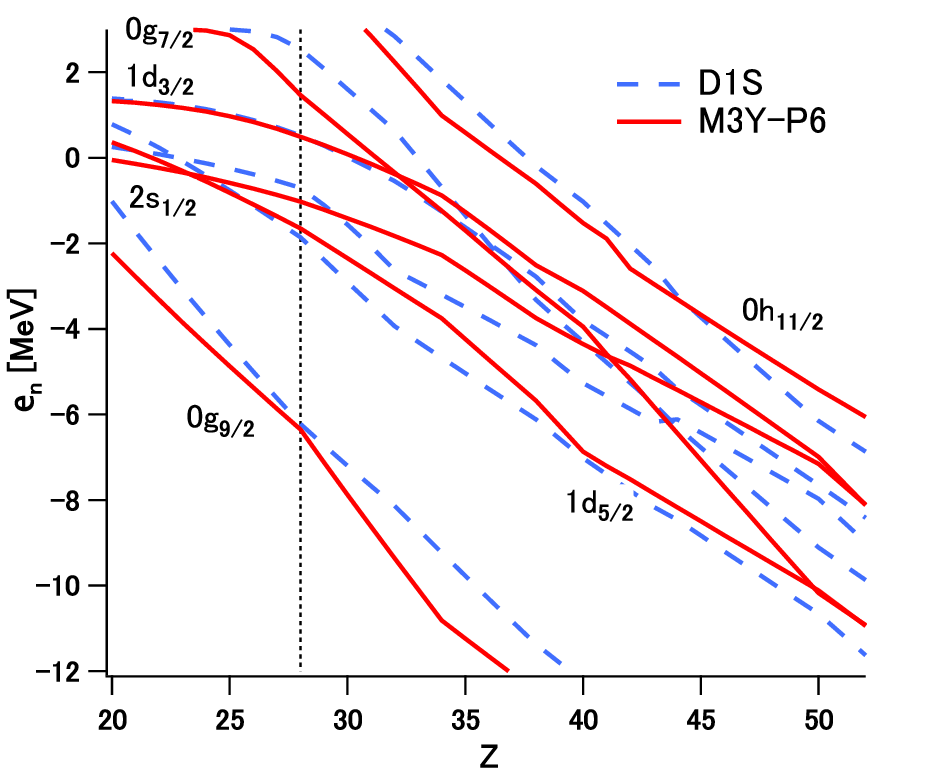}\quad
\raisebox{5mm}{\includegraphics[scale=0.5]{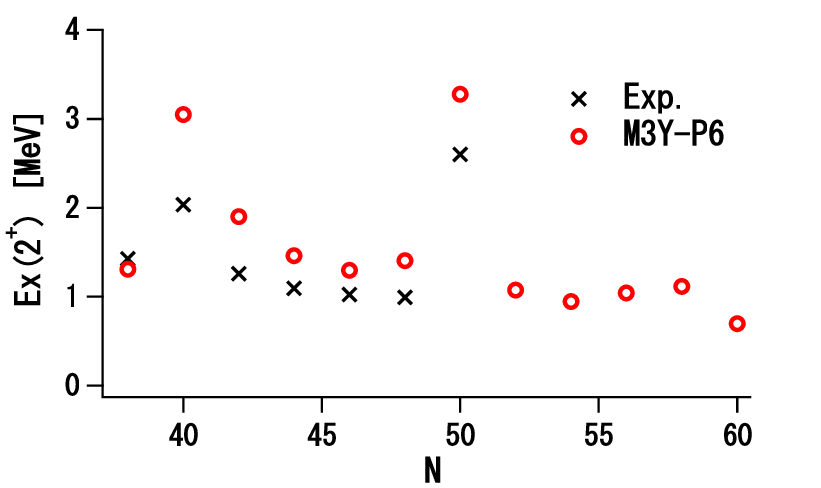}}
\caption{Left: Neutron s.p. levels in $N=58$ nuclei
  given by the HF calculations with M3Y-P6 (red solid line)
  and D1S (blue dashed line).
  Right: $E_x(2^+)$ of neutron-rich Ni nuclei
  predicted by the HFB\,+\,QRPA calculations with M3Y-P6 (red open circles),
  in comparison with the available experimental data (crosses)~\cite{NuDat}.}
  \label{fig:Ni86-magicity}
\end{center}
\end{figure}

The SCMF calculations with M3Y-P6 have been implemented
using the \texttt{MFGSB} computer code~\cite{mfgsb},
which can describe halos by superposing Gaussians~\cite{NS02,Nak06,Nak08}.
The single-particle (s.p.) levels
predicted by the spherical Hartree-Fock (HF) calculations for $N=58$
are shown in the left panel of Fig.~\ref{fig:Ni86-magicity}.
The s.p. levels of the Gogny-D1S interaction~\cite{D1S} are also presented.
We see a gap above the $n2s_{1/2}$ level,
indicating the magic or submagic nature of $N=58$ near Ni.
The highest occupied neutrons,
which occupy the $2s_{1/2}$ orbital,
are barely bound with $e_n(2s_{1/2})=-1.0\,\mathrm{MeV}$.
To further examine the $N=58$ magicity,
we have also carried out calculations
applying the quasiparticle random-phase approximation (QRPA)
on top of the Hartree-Fock-Bogoliubov (HFB) solutions
for neutron-rich Ni nuclei.
As depicted in the right panel of Fig.~\ref{fig:Ni86-magicity},
we observe an enhancement of $E_x(2^+)$ at $N=58$, \textit{i.e.}, $^{86}$Ni,
though it is suppressed due to the shallow binding of the last neutron.

\begin{wrapfigure}{r}{8cm}
\includegraphics[scale=0.5]{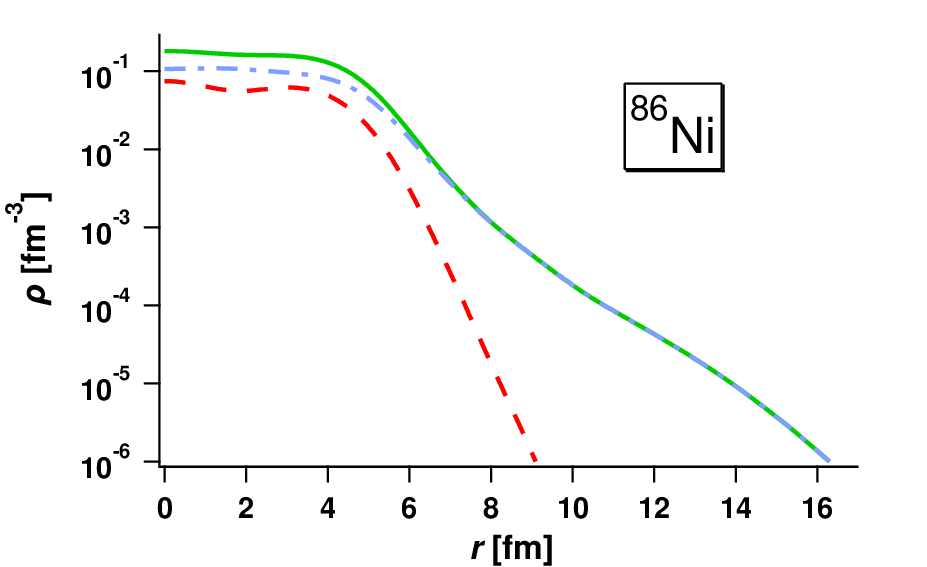}
\caption{Density distribution of $^{86}$Ni,
  predicted by the spherical HF calculation with M3Y-P6;
  $\rho_p(r)$ (red dashed line), $\rho_n(r)$ (blue dot-dashed line),
  and $\rho(r)=\rho_p(r)+\rho_n(r)$ (green solid line).
  Quoted from Ref.~\cite{IN25}.}\label{fig:Ni86_dns}
\end{wrapfigure}

Because of the shallow binding of the highest occupied neutrons,
it is expected that the doubly magic $^{86}$Ni may form a neutron halo.
This is confirmed by the predicted density distribution,
as is displayed in Fig.~\ref{fig:Ni86_dns}.

\section{Folding potential for halo nuclei --- $^{86}$Ni case}

\begin{wrapfigure}{l}{8cm}
\includegraphics[scale=0.6]{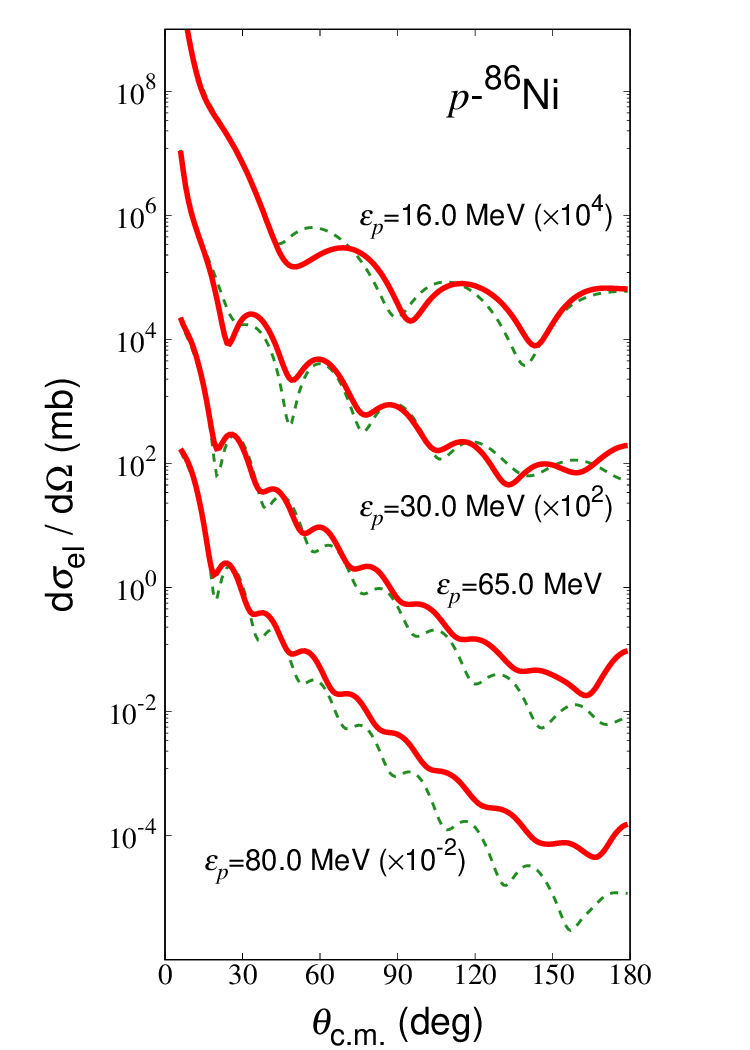}
\caption{Differential cross section of $p$-$^{86}$Ni elastic scattering
  $d\sigma_\mathrm{el}/d\Omega$ at several incident energies.
  Predictions of the SFP with M3Y-P6
  complemented with the KD imaginary potential (red solid line),
  and those of the fully empirical KD potential (green short-dashed line).}
\label{fig:DCS_Ni86}
\end{wrapfigure}

Using the SCMF wave function of $^{86}$Ni,
we compute the single-folding potential (SFP)
for $p$-$^{86}$Ni scattering.
In a fully consistent manner,
its real part is obtained from M3Y-P6 without local approximations.
The necessary formulae have been given in Ref.~\cite{NI24}.
It is noted that, owing to the non-locality of the potential,
this SFP (real part) has no energy-dependence.
To apply it to physical observables,
we need an imaginary part,
the derivation of which from the nucleonic interaction is beyond SCMF.
The imaginary part is complemented
by the Koning-Delaroche empirical potential~\cite{KD03}
for the sake of simplicity.
This approach has been confirmed to work well
at $\epsilon_p\lesssim 80^,\mathrm{MeV}$
for stable doubly-magic nuclei, from $^{16}$O to $^{208}$Pb~\cite{IN25,NI24}.
Scattering observables are calculated
via the \texttt{SIDES} computer code~\cite{SIDES}.
The results for the differential cross sections
are depicted in Fig.~\ref{fig:DCS_Ni86}.
They are compared with the results of the empirical potential,
employing the potential of Ref.~\cite{KD03}
both for the real and imaginary parts.

For stable nuclei,
the differential cross sections are close at low energies
between the SFP and the empirical potential~\cite{NI24}.
In contrast,
we find differences in the positions of peaks and dips for $^{86}$Ni.
These differences may be attributed partly to its large radius
due to the halo,
which is not taken into account in the empirical potential.
However, additional effects originating from broad density distribution
or non-locality in the SFP are indicated
by the dip and peak positions at $16\,\mathrm{MeV}$
and the structure around $20^\circ$ at $30\,\mathrm{MeV}$.
In Ref.~\cite{IN25},
the current SFP results without local approximations
have been shown to make a visible difference
from those obtained after the BR-LA.
The BR-LA on the LS channels, as well as on the central channels,
influences the results,
and the precision of the BR-LA is worse for $^{86}$Ni
than for stable nuclei.
Moreover, for $^{86}$Ni, the oscillating structure in the cross section
at $16\,\mathrm{MeV}$
is weakened in the BR-LA results,
unlike those of the full SFP and empirical potential.
The expansion on which the BR-LA is based leaves a room to be reexamined.
Notice that coupling to the continuum might not be included
sufficiently even in the present SFP.
If the current results or some extended ones are
compared with experimental data in the future,
it may provide insight into the influence of the continuum.

\clearpage
\section{Summary}

We have predicted and discussed,
based on SCMF calculations with the semi-realistic interaction M3Y-P6,
that $N=58$ could be magic near Ni
with the highest occupied neutron orbital being $2s_{1/2}$.
Thereby, $^{86}$Ni may be a unique nucleus,
begin doubly magic nucleus and having a neutron halo simultaneously.
If so, its relatively simple wave function
provides us with a good test case for, \textit{e.g.},
investigating optical potentials in halo nuclei.
Applying an approach consistent with the SCMF calculation
for the target nucleus,
which is a successful method for stable nuclei,
we calculate the (real part of) folding potential
for $p$-$^{86}$Ni scattering.
By comparing the results with those of the empirical potential,
we have pointed out the influence of the halo
and cast doubt on the conventional local approximation.

\section*{Acknowledgment}

This work is supported by MEXT/JSPS KAKENHI, Grant Nos. JP24K07012,
JP25H00402, and JP26H01399.
The numerical calculations were performed
using Oakforest PACS and Wisteria/BDEC, under the MCRP Program in CCS,
University of Tsukuba,
SQUID at D3 Center, The University of Osaka,
Yukawa-21 at YITP, Kyoto University,
and HITAC SR24000 at CUDTEC, Chiba University, in part.


\begin{thebibliography}{9}
\bibitem{Tan85} I. Tanihata, \textit{et al.},
  Phys. Rev. Lett. \textbf{55}, 2676 (1985).
\bibitem{Oza00} A. Ozawa, \textit{et al.},
  Phys. Rev. Lett. \textbf{84}, 5493 (2000).
\bibitem{Nak10} H. Nakada, Phys. Rev. C \textbf{81}, 051302(R) (2010).
\bibitem{IN25} K. Ishida, H. Nakada,
  Phys. Rev. C \textbf{111}, 044610 (2025).
\bibitem{Nak20} H. Nakada, Int. J. Mod. Phys. E \textbf{29}, 1930008 (2020).
\bibitem{Rii94} K. Riisager, Rev. Mod. Phys. \textbf{66}, 1105 (1994).
\bibitem{Tan13} I. Tanihata, H. Savajols, R. Kanungo,
  Prog. Part. Nucl. Phys. \textbf{68}, 215 (2013).
\bibitem{Ham17} I. Hamamoto, Phys. Rev. C \textbf{95} 044325 (2017).
\bibitem{NT18} H. Nakada, K. Takayama,
  Phys. Rev. C \textbf{98}, 011301(R) (2018).
\bibitem{BDP00} K. Bennaceur, J. Dobaczewski and M. P{\l}oszajczak,
  Phys. Lett. B \textbf{496}, 154 (2000).
\bibitem{Ots05} T.~Otsuka, \textit{et al.},
  Phys. Rev. Lett. \textbf{95}, 232502 (2005).
\bibitem{NS14} H. Nakada, K. Sugiura,
  Prog. Theor. Exp. Phys. \textbf{2014}, 033D02.
\bibitem{Nak13} H. Nakada, Phys. Rev. C \textbf{87}, 014336 (2013).
\bibitem{SNM16} Y. Suzuki, H. Nakada, S. Miyahara,
  Phys. Rev. C \textbf{94}, 024343 (2016).
\bibitem{MN18} S. Miyahara, H. Nakada,
  Phys. Rev. C \textbf{98} 064318 (2018).
\bibitem{CH89} R.L.~Varner, \textit{et al.},
  Phys. Rep. \textbf{201}, 57 (1991).
\bibitem{KD03} A.J.~Koning, J.P.~Delaroche,
  Nucl. Phys. A \textbf{713}, 231 (2003).
\bibitem{SL79} G.R.~Satchler and W.G.~Love,
  Phys. Rep. \textbf{55}, 183 (1979).
\bibitem{BR-LA} F.A.~Brieva and J.R.~Rook,
  Nucl. Phys. A \textbf{291}, 299 (1977);
  \textit{ibid.} \textbf{291}, 317 (1977);
  \textit{ibid.} \textbf{297}, 206 (1978).
\bibitem{mfgsb} H. Nakada, Chiba University Repository,
  DOI: 10.20776/900123722
  (https://opac.ll.chiba-u.jp/da/curator/900123722/?lang=1).
\bibitem{NS02} H. Nakada and M. Sato, Nucl. Phys. A \textbf{699}, 511 (2002);
  \textit{ibid.} \textbf{714}, 696 (2003).
\bibitem{Nak06} H. Nakada, Nucl. Phys. A \textbf{764}, 117 (2006);
  \textit{ibid.} \textbf{801}, 169 (2008).
\bibitem{Nak08} H. Nakada, Nucl. Phys. A \textbf{808}, 47 (2008).
\bibitem{D1S} J.F. Berger, M. Girod and D. Gogny,
  Comp. Phys. Comm. \textbf{63}, 365 (1991).
\bibitem{NuDat} http://www.nndc.bnl.gov/nudat3/.
\bibitem{NI24} H. Nakada, K. Ishida,
  Phys. Rev. C \textbf{109}, 044614 (2024).
\bibitem{SIDES} G.~Blanchon, \textit{et al.},
  Comp. Phys. Comm. \textbf{254}, 107340 (2020).
\end{thebibliography}
\end{document}